\documentclass[prl,twocolumn,superscriptaddress,aps]{revtex4-1}
\usepackage{amssymb,graphics,graphicx}
\usepackage{latexsym}
\usepackage{amsfonts}
\usepackage{amsmath}
\usepackage{float}

\newcommand{\be}{\begin{equation}}
\newcommand{\ee}{\end{equation}}
\newcommand{\bea}{\begin{eqnarray}}
\newcommand{\eea}{\end{eqnarray}}
\newcommand{\bma}{\begin{matrix}}
\newcommand{\ema}{\end{matrix}}

\newcommand{\bml}{\begin{mathletters}}
\newcommand{\eml}{\end{mathletters}}
\newcommand{\bes}{\begin{subequations}}
\newcommand{\ees}{\end{subequations}}

\newcommand{\bi}{\begin{itemize}}
\newcommand{\ei}{\end{itemize}}
\newcommand{\gev}{~{\rm GeV}}
\newcommand{\tev}{~{\rm TeV}}
\newcommand{\mev}{~{\rm MeV}}

\begin{document}
\title{Luminogenesis from Inflationary Dark Matter}
\author{Paul H. Frampton}
\email{paul.h.frampton@gmail.com}
\affiliation{Department of Physics and Astronomy, 
University of North Carolina, Chapel Hill, NC 27599-3255, USA}
\author{Pham Q. Hung}
\email{pqh@virginia.edu}
\affiliation{Department of Physics, University of Virginia,
Charlottesville, VA 22904-4714, USA;\\
Center for Theoretical and Computational Physics, 
Hue University College of Education, Hue, Vietnam}

\date{\today}

\begin{abstract}
A cosmological model is introduced in which dark matter plays
a dominant role throughout the history of the universe, and is
the only matter present for temperatures 
$T \gtrsim T_{15} \sim 10^9$ GeV.
The gauge group is $SU(3) \times SU(6)_{DM} \times U(1)_Y$
and unifies in such a way that luminous matter is 
generated at $T \sim T_{15}$
with the correct amount and eventual asymmetry. Construction
of more highly sensitive direct
detectors of dark matter {\it e.g.} XENON1T is encouraged. 
We offer a new explanation of why grand unification
theories involving only luminous matter may be fatally flawed.
\end{abstract}

\maketitle
\section{Introduction}
From the latest Planck data \cite{Planck},
it is believed that dark matter makes up for 
around 27 \% of the energy density of the universe 
while luminous matter makes up around 4.5 \%, 
the rest being in the form of Dark Energy. Furthermore, structure 
formation in the Universe is generally believed to be driven 
by Dark Matter. 

The fact that Dark Matter constitutes the 
dominant form of matter in the present time is remarkable. 
If it is so now, it is extremely reasonable to think that it  
was perhaps also dominant in the early universe. 
And perhaps, it was the only matter that existed in the 
very early universe. The generation of luminous matter 
would come about when a fraction of dark matter converted 
into luminous matter. The size of that fraction would 
depend on the efficiency of the conversion process. 
The temperature (or energy) where this conversion took 
place would naturally depend on the dark matter mass(es) 
and its conversion efficiency would depend on another mass 
scale which governs the strength of the interaction.
It is necessary that dark matter is unified with luminous 
matter in the underlying gauge theory. We require that 
dark matter interacts with luminous matter strongly enough 
to deplete the initial amount of dark antimatter 
(and hence dark anti-matter) leaving an excess in dark matter 
which leads eventually to an excess in luminous matter. 
However, it should at the same time be weakly interacting 
enough to escape direct detection at the present sensitivity.
For a recent summary of the status of non-WIMP Dark Matter, one can consult \cite{kusenko}.

\bigskip

\noindent
Our model is based partially on the scenario 
presented in \cite{paulpqdm} 
where it was proposed that dark matter particles 
(fermions) come in {\em two} species: $\chi_l$ and $\chi_q$ 
which transform under the product of a dark matter gauge group with
the standard model (SM), 
namely $SU(4)_{DM} \times SU(3)_c \times SU(2)_L \times U(1)_Y$, 
as $(4,1,1,0)_{L,R}$ for $\chi_l$  and $(4,3,1,0)_{L,R}$ 
for $\chi_q$. The particles 
$\chi_l$ and $\chi_q$ will be referred to as  
"leptonic" (color singlet) and "baryonic" (color triplet) 
dark matter respectively. In the present model
the dark matter carries no SM 
quantum numbers so that only the field $\chi_l$ is involved.
A similarly motivated but technically completely different 
model has been recently and presciently built in \cite{nath}.

Our model treats the symmetric and asymmetric dark matter differently
from the conventional approaches. Only
the asymmetric part of the dark matter survives to the present
time: all the symmetric part annihilates long ago. Some
14 \% of the asymmetric dark matter transmutes into luminous
matter via our luminogenesis mechanism. In particular, we do
not need the conventional WIMP annihilation cross section value
to obtain the correct relic density which is obtained correctly
by another mechanism. Again, we would like to stress that the usual mechanism
to determine the relic density which relies on the annihilation cross section, even in the presence of an asymmetric part
of DM \cite{asym}, does not apply to our model. This point will be clarified further below.

\section{Inflationary Dark Matter}

\subsection{Dark Matter in SU(6)}
If the dark matter field $\chi_l$ is to be a singlet under the 
SM gauge group and if it were to be unified with luminous matter, 
its own gauge group $G_{DM}$ (if there were one) should be 
embedded in a larger dark unification group
$G_{DUT}$ which contains the SM group $G_{SM}$, 
namely $G_{DUT} \rightarrow G_{DM} \times G_{SM}$.  
This unified group would be one on which inflation is based such that
the inflaton will decay into dark matter during the reheating process.

\bigskip

\noindent
We use the model proposed to unify dark matter with luminous 
matter in \cite{paulpqdm}. The unification of the two 
sectors proceeds via the embedding of  $SU(2)_L$ into a unifying 
group $SU(n+2)$ with the following breaking path 
$SU(n+2) \times U(1)_Y \rightarrow 
SU(n)_{DM} \times SU(2)_L \times U(1)_{DM} \times U(1)_Y$. 
Including QCD, the unifying group would be 
$SU(3)_C \times SU(n+2) \times U(1)_Y$ 
("unifying" solely in the sense of dark and luminous matter 
unification and not in the usual sense of gauge unification). 
In \cite{paulpqdm}, arguments were given for the selection of 
the preferred value
$n=4$ for the dark matter gauge group and our final choice is
\be
\label{group}
SU(3)_C \times SU(6) \times U(1)_Y
\ee 
\noindent
with $SU(6)$ subsequently breaking according to
\be
SU(6)  \rightarrow  
SU(4)_{DM}  \times U(1)_{DM} \times SU(2)_L \, .
\label{su6}
\ee
Here $G_{DUT}$ is $SU(6)$ and $G_{DM}$ is $SU(4)$. 
It is convenient to show the various useful representations 
of $SU(6) \supset SU(4)_{DM} \times SU(2)_L \times U(1)_{DM}$.
\bea
\label{rep}
6& =& (1,2)_2 + (4,1)_{-1} \nonumber \\
20 &=& (4,1)_3 + (4^\ast,1)_{-3} + (6,2)_0 \nonumber \\
35 &=& (1,1)_0 + (15,1)_0 +(1,3)_0 + (4,2)_{-3} + (4^\ast ,2)_3 \nonumber \\
\eea
where $U(1)_{DM}$ quantum numbers are indicated by subscripts. 
Note that $U(1)_{DM}$ will be spontaneously broken at a 
scale $\Lambda_{DM}$ which will be constrained by experimental 
direct detection limits. The associated massive gauge boson, 
$\gamma_{DM}$, is the oft-discussed "dark photon". We shall 
come back to this important point below. 

At a scale $\Lambda_4$, $SU(4)_{DM}$ will become confining 
and DM hadrons form as has been discussed in \cite{paulpqdm}. 
The fact that our model contains strongly self-interacting 
dark matter is an interesting feature which might resolve 
the well-known $\Lambda$CDM problems \cite{CDMissues}
of dwarf galaxy structures
and of dark matter cusps at the centers of galaxies. This discussion is presented below.

From Eq. (\ref{rep}), the representations that contain 
singlets under the SM $SU(2)_L$ gauge group are 
$\underline{6}$ and $\underline{20}$. These are the 
representations that could contain the desired dark 
matter particles, namely $(4,1)$ which appears in 
both $\underline{6}$ and $\underline{20}$. To see 
where the dark matter belongs, it is important to 
classify the fermion representations using Eq. (\ref{rep}).

As discussed above, our unified gauge group is 
$SU(3)_C \times SU(6) \times U(1)_Y$. The fermion 
representations are required to be anomaly-free. 
Representations containing the left-handed SM 
quark and lepton doublets are respectively 
$(3,6,Y_{6q} /2)_L$ and $(1,6,Y_{6l} /2)_L$ 
where $Y_{6q,l} /2$ are the $U(1)_Y$ quantum numbers 
of the quarks and leptons respectively. In addition, 
the $SU(2)_L$ quark and lepton singlets are written 
as $(3,1,(Y_{u} /2, Y_{d} /2)_R$ and  
$(1,1,Y_{l} /2)_R$ respectively. The $U(1)_Y$ quantum 
numbers are, as usual, $Y_{6q} /2= 1/6$ 
and $Y_{6l} /2=-1/2$ for $SU(6)$ non-singlets 
and $Y_{u} /2=2/3$, $Y_{d} /2=-1/3$, and $Y_{l} /2=-1$.
Since $\underline{3}$ and $\underline{6}$ 
are complex representations, the minimal anomaly-free 
representations are given by
\bea
\label{rep2}
&&(3,6,Y_{6q} /2)_{L,R} + (1,6,Y_{6l} /2)_{L,R} 
+ (3,1,(Y_{6u} /2, Y_{6d} /2)_{R,L} \nonumber \\
&&+ (1,1,Y_{l} /2)_{R,L} \, .
\eea
\noindent
As we have mentioned in \cite{paulpqdm}, the right-handed 
quark and lepton doublets, 
$(3,6,Y_{6q} /2)_{R} + (1,6,Y_{6l} /2)_{R} $, 
and left-handed singlets, 
$(3,1,(Y_{6u} /2, Y_{6d} /2)_{L} + (1,1,Y_{6l} /2)_{L}$,  
are in fact the mirror fermions (distinct from SM fermions) 
of the model of electroweak-scale right-handed neutrinos in
\cite{hung}. The details of how right-handed neutrinos, which 
are members of doublets along with their mirror charged lepton 
partners, can acquire electroweak-scale mass can be found 
in \cite{hung}, where arguments were given for assigning 
the mirror sector a global symmetry. As discussed in \cite{hung}, this mirror sector can be tested experimentally
at the LHC by looking for lepton-number violating processes through the production of electroweak-scale right-handed
neutrinos. This model fits rather well the electroweak precision parameter constraints as shown in \cite{hung2}.
In light of the 
newly-discovered SM-like 126 GeV scalar, an extension of 
the model of \cite{hung} to endow {\em separately} the SM sector and the 
mirror sector with a global symmetry is needed \cite{ajinkya}: 
a global $U(1)_{SM} \times U(1)_{MF}$ ("MF" stands for the mirror sector) is imposed, with two Higgs doublets, one for each sector. As shown in \cite{ajinkya},
a small mixing allowed by the present data on the 126-GeV scalar breaks explicitly this global $U(1)_{SM} \times U(1)_{MF}$ symmetry
with an interesting indirect implication on luminogenesis as we will discuss below.

From Eqs.(\ref{rep},\ref{rep2}), one can see that the 
$SU(2)_L$-singlet and $SU(4)$-non-singlet particles 
transform under 
$SU(3) \times SU(4)_{DM} \times SU(2)_L \times U(1)_Y 
\times U(1)_{DM} $ as (for both left and right-handed fermions)
\be
\label{rep3}
(1, 4, 1, -\frac{1}{2})_{-1} + (3, 4, 1, \frac{1}{6})_{-1} \, ,
\ee
where the subscripts $U(1)_{DM}$ quantum numbers. 
It is clear from (\ref{rep3}) that these particles 
which belong to the $6$ of $SU(6)$ {\em cannot} be 
candidates for dark matter since they carry $U(1)_Y$ 
quantum numbers and are therefore electrically charged. 
In fact, the color-singlet and colored particles carry 
charges $\pm 1/2$ and $\pm 1/6$ respectively.
A suitable representation which is color-singlet and carries 
no $U(1)_Y$ quantum number is the following real representation:
\be
\label{rep4}
(1, 20, 0) = (1,4,1,0)_3 + (1, 4^\ast,1,0)_{-3} + (1, 6, 2, 0)_0 \, ,
\ee
where the right-hand side represents decompositions under 
$SU(3) \times SU(4) \times SU(2)_L \times U(1)_Y \times U(1)_{DM} $. 
One notices that $(1,4,1,0)_3 + (1, 4^\ast,1,0)_{-3}$ 
are {\em inert} under the SM gauge group 
$SU(3) \times SU(2)_L \times U(1)_Y$ but not under 
$U(1)_{DM} $. These particle are the dark matter in our model: 
note that, when one represents fermions in terms of left-handed 
Weyl fields, we have $\chi_{L,R}= (1, 4^\ast,1,0)_{-3}$ 
and $\chi^{c}_{L} = \sigma_2 \chi^{\ast}_{R}=(1,4,1,0)_3$.  
How the dark matter candidates are produced in the early 
universe and how luminogenesis, the generation of luminous
matter from dark matter, occurs will be discussed in the next 
two subsections.

For the sake of clarity, two tables are given below in order to list the different particle contents of the model.
 \begin{table}[h]
 \label{Tab1}
 \begin{tabular}{|l|l|lr|||} \hline
 $SU(6)$ & $SU(4) \times SU(2) \times U(1)_{DM}$ \\ \hline
 ${\bf 6}$ & ${\bf (1,2)_2 + (4,1)_{-1}}$  \\
 $ {\bf 20}$ & ${\bf (4,1)_3 + (4^\ast , 1)_{-3} + (6,2)_{0} }$ \\
 ${\bf 35}$ & ${\bf (1,1)_0 + (15,1)_0 + (1,3)_0 +(4,2)_{-3}}$  \\ 
 & ${\bf  + (4^\ast , 2)_3}$ \\  \hline
 \end{tabular}
 \caption{The ${\bf (1,2)_2}$'s represent luminous matter while ${\bf (4,1)_3 + (4^\ast , 1)_{-3}}$ represent dark matter}
 \end{table}
 \begin{table}[h]
 \label{Tab2}
 \begin{tabular}{|l|l|lr|||} \hline
  & $SU(3)_c \times SU(6) \times U(1)_{Y}$ \\ \hline
 R $\supset$ SM fermions & ${\bf (3,6, 1/6)_L + (1,6, -1/2)_L }$  \\
 &${\bf + (3,1,2/3)_R + (3,1,-1/3)_R }$ \\ 
 &${\bf + (1,1,-1)_R}$ \\
 \hline
 R $\supset$ Mirror fermions &  ${\bf (3,6, 1/6)_R + (1,6, -1/2)_R }$  \\
 &${\bf + (3,1,2/3)_L + (3,1,-1/3)_L }$ \\ 
 &${\bf + (1,1,-1)_L}$ \\
 \hline
 R $\supset$ dark matter fermions & ${\bf (1,20,0)}$ \\  \hline
 \end{tabular}
 \caption{R denotes representation. SM left-handed doublets and right-handed singlets are parts of the first entry, Mirror right-handed doublets and left-handed singlets are parts of the second entry, and dark matter left and right-handed fermions belong to the last entry.}
 \end{table}

\subsection{Dark matter genesis}

We assume that the potential for the the adjoint scalar field 
$\underline{35}$ of $SU(6)$ is {\em sufficiently flat} so as 
to generate sufficient inflation at the scale of $SU(6)$ breaking. 
It is beyond the scope of this article to treat this aspect of 
inflation and we will restrict ourselves to its group theoretic
aspects. The inflaton field is the $\phi_{inf} =(1,1,1,0)_0$ of
\bea
\label{adj}
(1,35,0)& =& (1,1,1,0)_0 + (1,15,1,0)_0 +(1,1,3,0)_0  \nonumber \\
&&+ (1,4,2,0)_{-3} + (1,4^\ast ,2,0)_3 \,,
\eea
where the right-hand-side shows the transformation under 
$SU(3) \times SU(4) \times SU(2)_L \times U(1)_Y \times U(1)_{DM} $.  
The fermions that can couple to the adjoint scalar will come from 
$ 20 \times 20 = 1_s + 35_a + 175_s + 189_a$  and 
$6 \times \bar{6} = 1 + 35$.  Denoting $(1, 20, 0)$ by 
$\Psi_{20}$ and $(1,35,0)$ by $\phi_{35}$, one can write the 
following coupling
\be
\label{yuk20}
g_{20} \, \Psi_{20}^{T} \sigma_2 \Psi_{20} \, \phi_{35} \, .
\ee
From Eq.~(\ref{yuk20}), one can deduce the 
coupling of the inflaton to dark matter
\be
\label{infdm}
g_{20} \, \chi_{L}^{T} \sigma_2 \chi^{c}_{L} \phi_{inf} \, .
\ee

\noindent
Since $\psi^{c}_L = \sigma_2 \psi^{\ast}_R$, it is clear 
that $\bar{6} \sim \psi^{c}_{6,L} $ comes from mirror 
fermions and $6 \sim \psi_{6,L}$ contains SM fermions. 
As a result, a coupling such as 
$\psi^{c}_{6,L} \sigma_2 \psi_{6,L} \, \phi_{35}$ is forbidden 
at tree-level by the $U(1)_{SM} \times U(1)_{mirror}$ symmetry. 
The inflaton will decay mainly into dark matter while its 
decay into luminous matter will be highly suppressed by the 
aforementioned symmetry.
Another interesting point that one could point out here is 
quantum fluctuations during the inflationary period can create 
seeds of structure formation but in our scenario, it is 
structures of dark matter that were formed first. This is 
actually the current view of structures in the universe. 
In our model, structures involving luminous matter came only 
later when approximately 14 \% of dark matter is converted 
into luminous matter.

The next section discusses this conversion of some of the 
dark matter energy density into luminous matter, a process 
we call {\em luminogenesis}. In particular, we will present arguments showing that
the symmetric part of DM annihilates "almost completely" into the symmetric part of luminous
matter which, in turn, transforms into radiation. The asymmetric part of DM transforms a small part
of its number density into the asymmetric part of luminous matter through an {\em entirely different} mechanism
and is unrelated to and unconstrained by the annihilation cross section used in the symmetric part.

\subsection{Luminogenesis}

This section deals with the {\em fate} of the asymmetric and symmetric parts of DM. A few words concerning a possible origin of the excess of dark matter over anti-dark matter (the asymmetric part) is in order here.

First, we will assume that there is a global $U(1)_{\chi}$ symmetry for dark matter. The interactions involving the gauge bosons of the coset group $SU(6)/SU(4) \times SU(2) \times U(1)_{DM}$ (similar to X and Y gauge bosons of $SU(5)$) will explicitly violate the $U(1)_{\chi}$ symmetry and their decays involving the interferences between the tree-level and one-loop diagrams will ultimately generate a net DM number assuming the presence of CP violation in the DM sector. This problem will be treated in a separate paper. For the present purpose, we will assume that the asymmetric part is generated by the aforementioned mechanism.

Let us denote the asymmetric number density by $\Delta n_{\chi} = n_{\chi} - n_{\bar{\chi}}$ and the symmetric number density simply by $n_{sym} = n_{\bar{\chi}}$ since the symmetric part is composed of an equal number of DM and anti-DM. It is assumed that $\Delta n_{\chi} \ll n_{sym}$. What we will show below will be that the symmetric part will annihilate through the massive dark photon of $U(1)_{DM}$ into the symmetric part of luminous matter which will eventually transform into radiation, leaving practically very little symmetric DM. As we will also show, this has {\em no bearing} on the relic density of DM and hence that of its luminous off-spring. A fraction ($\sim 14 \%$) of the asymmetric part of DM, $\Delta n_{\chi}$, is transmuted into the asymmetric part of luminous matter through an {\em entirely different} mechanism coming from an exchange of a massive scalar. This is {\em different} from the usual approach whereas the same annihilation process determines both the relic asymmetric and symmetric densities which are therefore intrinsically linked \cite{asym}.

The simplified discussion below will go as follows. There are two different interactions which operate on the total DM density $n_{tot} = n_{sym} + \Delta n_{\chi}$.
The exchange of the massive scalar (to be discussed next) affects both the symmetric and asymmetric parts. Constraints on the scalar mass and the Yukawa couplings are imposed in such a way that 14 \% of the DM (i.e. 14 \% of $n_{sym}$ and 14 \% of $\Delta n_{\chi}$) is converted into luminous matter. The decoupling of the scalar exchange interaction should happen soon after DM becomes non relativistic. At the same time the symmetric part will annihilate into luminous matter via the massive dark photon. This will reduce the symmetric part to slightly below 86 \% by the time of the scalar exchange decoupling. For our simplified discussion, we will ignore this difference. The 86 \% of the symmetric DM will continue to annihilate via the massive dark photon until very little is left as we will show below. What is {\em unaffected} by the $U(1)_{DM}$ interactions is the 86 \% of the asymmetric part and the 14 \% of the asymmetric luminous matter. This is the essence of our luminogenesis. It goes without saying that a more accurate analysis involving a numerical study of the Boltzmann equation as well as the inclusion of chemical equilibrium is needed and this will be carried out in a future work. For this paper, we present a simplified discussion in order to lay out the essence of our model.




\bigskip

\noindent


\bigskip
\noindent
{\bf I}. {\bf The conversion of the symmetric and asymmetric parts via a scalar exchange:}
\bigskip

We present in this section the main mechanism for luminogenesis: The conversion of a small fraction of DM into luminous matter by the exchange of a heavy scalar. We will show the constraints on the Yukawa couplings and the mass of the heavy scalar coming from the requirement that 14 \% of DM is converted into luminous matter. In the next section, we will discuss how most of the 86 \% of the symmetric part of DM annihilates via a massive dark photon into the symmetric luminous matter which eventually transforms into radiation.  The 86 \% of asymmetric DM and the 14 \% of asymmetric luminous matter are unaffected by this annihilation process.
\bigskip

Ia) {\bf The interaction Lagrangian and related features:}

Since the dark and luminous sectors 
belong to different representations of $SU(6)$, the conversion can be
achieved only through a coupling of the dark and luminous 
sectors with a scalar field.
Since $20 \times \bar{6} = 15 + 105$ and 
$20 \times 6 = \bar{15} + \bar{105}$ ($20$ is real), 
the appropriate scalars transform as $\bar{15}$ and $15$ respectively. 
We denote these scalars as $\Phi_{15}^{(L)} (1/2)$ and 
$\Phi_{\bar{15}}^{(R)} (-1/2)$ where $\pm 1/2$ denotes the 
$U(1)_Y$ quantum number. We have the following Yukawa couplings
\be
\label{DMLUM}
g_{6L} \, \Psi_{20}^{T} \sigma_2 \psi_{6,L} \, \Phi_{15}^{(L)} + g_{6R} \, \Psi_{20}^{T} \sigma_2 \psi^{c}_{6,L} \, \Phi_{\bar{15}}^{(R)} \, ,
\ee
with $\Phi_{15}^{(L)}$ and $\Phi_{\bar{15}}^{(R)}$ carrying  
appropriate global $U(1)_{SM}$ and $U(1)_{mirror}$ quantum numbers 
respectively. 
A mass mixing between $\Phi_{15}^{(L)}$ and 
$\Phi_{\bar{15}}^{(R)}$ will break the global 
$U(1)_{SM} \times U(1)_{MF}$ symmetry and thus
allows for the following conversion process to occur: 
$\chi_L + \chi_R \rightarrow l_L + l^{M}_R $ 
where $l_L $ and $l^{M}_R$ refer to SM and mirror leptons 
respectively as in \cite{hung}. The same goes for the anti-DM particles. This process can be represented 
by the following effective Lagrangian
\be
\label{eff}
\frac{g_{6}^2}{M_{15}^2} \, (\chi^{T}_{L} \sigma_2 l_L)\, (\chi^{c,T}_{L} \sigma_2 l^{M,c}_{L}) + H.c. \, ,
\ee
where $l^{M,c}_{L} = \sigma_{2}  l^{M \ast}_{R}$.  
The various mixing coefficients are embedded in the prefactor 
of Eq.~(\ref{eff}). How effective the conversion of dark matter 
into luminous SM {\em and} mirror leptons as represented by 
Eq.~(\ref{eff}) will depend on this prefactor, especially
the luminogenesis scale $M_{15}$ which in the next subsection we shall estimate
to be $M_{15} \sim 10^9$ GeV. Notice that mixing between $\Phi_{15}^{(L)}$ and 
$\Phi_{\bar{15}}^{(R)}$ has to be sufficiently small so that only a small fraction of DM is converted into luminous matter, namely $\sim$ 14 \%.
As we have mentioned above, it is interesting to note that an extended version of \cite{hung} to include two Higgs doublets (with similar
$U(1)_{SM} \times U(1)_{MF}$ assignments as $\Phi_{15}^{(L)}$ and 
$\Phi_{\bar{15}}^{(R)}$) in
order to describe the 126-GeV Higgs-like object also requires the mixing between these two doublets to be small as constrained by the data
\cite{ajinkya}.

At this point we would like to point out an important fact that comes out of the model of \cite{hung}, namely the decay of the
mirror lepton $l^{M}_R$ into a SM lepton $l_L$. This decay proceeds through a Yukawa interaction
\be
\label{mirrortol}
g_{sl} \, \bar{l}_{L}\, \phi_S \, l^{M}_{R} + H.c. \, ,
\ee
where $\phi_S$ is the SM-singlet Higgs field with $l_L$ and $l_{R}^{M}$ being SM left-handed and mirror right-handed doublets
respectively. Notice that under $U(1)_{MF}$, $l^{M}_{R} \rightarrow \exp (\imath \alpha_{MF}) l^{M}_{R}$ while under $U(1)_{SM}$, $l_{L} \rightarrow \exp (\imath \alpha_{SM}) l_{L} $. As a result, $\phi_{S} \rightarrow \exp (\imath (\alpha_{SM} - \alpha_{MF}))$. This point is explained in \cite{hung} for $U(1)_{MF}$ and extended to $U(1)_{SM} \times U(1)_{MF}$ in \cite{ajinkya}.

The physical reason for writing down Eq.~(\ref{mirrortol}) is given in \cite{hung} where a model for an electroweak-scale Majorana mass
for the right-handed neutrinos was constructed. The singlet Higgs boson $\phi_S$ is vey light ($\sim 100 keV$ or even $O(\mev)$), as discussed in \cite{hung}. (Notice that in \cite{hung}, $g_{sl} \, v_S \sim 10^{5} eV$ and for simplicity it was assumed that $g_{sl} \sim O(1)$ giving $v_{S} \sim 10^{5} eV$. However one can have $g_{sl} \sim 10^{-2}$ which gives $v_{S} \sim 10 \mev$ and hence a mass of $\phi_S$ in the MeV range. (Or it could even be in the tens of keVs.) Eq.~(\ref{mirrortol}) gives rise to the decay $l_{R}^{M} \rightarrow l_L + \phi_S$. (This includes decays of mirror charged and neutral leptons.)
Depending on the size of the coupling $g_{Sl}$, $l_{R}^{M} $ could be relatively long-lived with distinct signatures at the LHC. However, on a cosmic scale, the mirror leptons $l_{R}^{M}$ promptly decay into SM leptons. The end product of Eq.~(\ref{eff}) will be the conversion of a fraction of the DM particles into SM leptons only with no mirror leptons left. The details of this conversion process are important and will be treated elsewhere.

The conversion of the SM leptonic asymmetry to the baryonic asymmetry can proceed via the well-known sphaleron process \cite{yanagida}.

Although the fate of $\phi_S$ was discussed in \cite{hung}, it is useful to repeat it here. From \cite{hung}, one can obtain the interactions between $\phi_S$, $\nu_R$ (with $M_R \sim O(\Lambda_{EW})$), and $\nu_{L}$, as well as with $e^{M}_R$ and $e_L$ as $g_{Sl} \bar{\nu}_L \phi_{S} \nu_R + H. c.$ and $g_{Sl} \bar{e}_L \phi_{S} e^{M}_R + H. c.$, coming from $g_{Sl} \bar{l}_L \phi_{S} l_R + H. c.$. $\phi_S$ is in thermal equilibrium with luminous matter through the reactions $\phi_S + \phi_S^{\ast} \leftrightarrow \bar{\nu}_L + \nu_L$, $\phi_S + \nu_L  \leftrightarrow \phi_S + \nu_L$, $\phi_S + \phi_S^{\ast} \leftrightarrow \bar{e}_L + e_L$, $\phi_S + e_L  \leftrightarrow \phi_S + e_L$. To see their effects on BBN, we shall use the analysis of \cite{dreiner} for some particular range of values for $m_{\phi_S} $, namely in the tens of keVs. Without repeating what has been done in \cite{dreiner}, it is illuminating to stress that if there are light particles that couple to matter and if these light particles decouple close to the temperature where neutrinos decouple, they can be counted toward the effective number of neutrinos which may exceed the cosmological constraints of BBN. In a nutshell, if the rate is comparable to the weak interaction rate, the effective number of neutrinos might exceed the current bound. Notice that $C_{x} G_{x}$ of \cite{dreiner} is identified in the model of \cite{hung} and hereon as $\sim G_{\phi_S} = g_{sl}^2/(4\sqrt{2} M_R^2)$ where $M_R \sim O(\Lambda_{EW})$ is a typical mirror fermion mass. The constraint obtained by \cite{dreiner} by requiring the effective number of neutrinos $N_{eff} < 4$  with the neutrino decoupling temperature $T_{\nu} \sim 3\, \tev$, for a light complex scalar (our case), is then
\be
\label{phi}
G_{\phi_S}  \alt 4.1 \times 10^{-2} C_V \, G_F \,,
\ee
where $C_V \sim O(1)$ and $G_F = g^2/(4\sqrt{2} M_W^2) = 1.166 \times 10^{-5} \gev^{-2}$ is the Fermi constant. The constraint (\ref{phi}) can be easily satisfied for $g_{sl}^2 < 10^{-2} g^2$.  It is interesting to note that a very small value for $g_{sl}$ can lead to a long-lived mirror lepton and this could have interesting implications at the LHC \cite{hung} as well as facilities searching for rare decays such as $\mu \rightarrow e \gamma$ and $\tau \rightarrow \mu \gamma$ \cite{hung2}. A more comprehensive study of the cosmological implications of $\phi_S$ is beyond the scope of this paper and will be presented elsewhere.


We now proceed to discuss the conversion of part of the asymmetric DM into the asymmetric luminous matter.
\bigskip

Ib) {\bf Dark and Luminous matter densities:}

\bigskip

 In what follows we will present a simplified version of luminogenesis which captures the essence of the process. A more complete numerical analysis of the Boltzmann equation and chemical potential equilibrium will be presented elsewhere. 

Recall that $n_{tot} = n_{sym} + \Delta n_{\chi}$. We assume that the decoupling occurs fast enough so that the universe is still matter-dominated. The interaction (\ref{eff}) decouples from the DM when the interaction rate $\Gamma \approx (\alpha_{6}^2/ M^{2}_{15}) n_{tot} $ is less than the Hubble rate $H= \sqrt{8 \pi/3} (1/m_{pl}) \sqrt{\rho_m} = \sqrt{8 \pi/3} (1/m_{pl}) \sqrt{n_{tot} m_{\chi}}$. The total density at decoupling is obtained from $\Gamma = H$ giving
\be
\label{15}
n_{tot,D} \approx (\frac{8 \pi}{3})(\frac{1}{\alpha_{6}^4})(\frac{M_{15}}{m_{pl}})^2 m_{\chi} M^{2}_{15} \,.
\ee
Let us define
\be
\label{ratio1}
r= \frac{n_{tot,D}}{n_{tot,0}} \,.
\ee
where $n_{tot,0}$ is the number density at $T \sim m_{\chi}$ and is given by $n_{tot,0} \sim C m_{\chi}^3$. For the sake of estimate, we will not be very precise about the exact value of $C$.  We obtain
\be
\label{ratio2}
r \approx \tilde{C} (\frac{1}{\alpha_{6}^4})(\frac{M_{15}}{m_{pl}})^2 (\frac{M_{15}}{m_{\chi}})^2 \,,
\ee
where $\tilde{C} \sim O(10^2)$. Eq.~(\ref{ratio2}) tells us about the relationship between the coupling $\alpha_{6}$ and the mass $M_{15}$ for a given DM mass $m_{\chi}$ with $r$ set to $r= 86 \%$.  From Eq.~(\ref{eff}), one can see that the cross section would increase as $M_{15}$ decreases and if we want to keep $r$ fixed, one has to decrease $\alpha_{6}$ as well. For $r= 86 \%$ one can solve for $\alpha_6$ as a function of $M_{15}$ and $m_{\chi}$ and which goes like
\be
\label{alpha6}
\alpha_6 \approx 1.16 \tilde{C} \sqrt{\frac{M_{15}}{m_{pl}}} \sqrt{\frac{M_{15}}{m_{\chi}}} \,.
\ee
As an example, let us take $m_{\chi} \sim 1\, \tev$. One gets $\alpha_6 \sim 10^{-2}, 10^{-3}, 10^{-4}, 10^{-5}$ for $M_{15} \sim 10^{9}, 10^{8}, 10^{7}, 10^{6}$ GeV respectively. Notice that what we call by $\alpha_6$ is actually a quantity which contains various factors such as the square of the Yukawa couplings and, most importantly, the mixing angle between $\Phi_{15}^{(L)}$ and $\Phi_{\bar{15}}^{(R)}$ as discussed above. At this point, it is interesting to note that it is plausible that the mixing between $\Phi_{15}^{(L)}$ and $\Phi_{\bar{15}}^{(R)}$ is similar to that between the two Higgs doublets in an extension \cite{ajinkya} of \cite{hung}. This opens up the possibility that luminogenesis could be indirectly tested at the LHC.

The final accounting goes as follows. At decoupling, there are 86 \% asymmetric DM, 14 \% asymmetric luminous matter and the same goes for the symmetric parts. The 86 \% of symmetric DM will in turn annihilate into symmetric luminous matter and quickly gets depleted as we will show in the next section.

\bigskip

\noindent 
{\bf II}. {\bf The symmetric part:}

Since both dark and luminous matter carry nonzero $U(1)_{DM}$ quantum 
number, dark matter can annihilate via the $\gamma_{DM}$ massive gauge 
boson into particle-antiparticle pairs of the luminous sector. This can be represented by an effective interaction
\be
\label{gamma}
\frac{g^2}{M_{\gamma_{DM}}}(\bar{\chi}  \gamma_{\mu} \chi)(\bar{f} \gamma^{\mu} f) \,.
\ee

The amount of symmetric DM which remains after the above decoupling is actually lower than 86 \% due to annihilation via $\gamma_{DM}$. However, the number left over after the $U(1)_{DM}$ interaction decouples is so small that we will ignore this difference. The argument goes as follows.

Following a similar reasoning to the above analysis, we look for the number density of the symmetric part of DM at the time of decoupling i.e. its value when the interaction rate is equal to the Hubble rate. There is however a difference with the above analysis concerning the Hubble rate at decoupling. It is reasonable to assume that the temperature at which the $U(1)_{DM}$ interaction goes out of thermal equilibrium to be much lower than $m_{\chi}$. For example, the energy density ratio for $T= m_{\chi}/10$ is roughly $\rho_{\chi}/\rho_{R} < \exp(-10) \approx 4.5 \times 10^{-5}$, implying a radiation-dominated universe. The density of symmetric DM at decoupling is determined by
\be
\label{dark}
\frac{\alpha^2}{M_{\gamma_{DM}}^2}\, n_{sym,D} \approx \frac{T_{D}^2}{m_{pl}} \,,
\ee
where $D$ again stands for "decoupling". Again using $n_{tot,0} \sim n_{sym,0} \sim C m_{\chi}^3$, one obtains 
\be
\label{ratio3}
\frac{n_{sym,D}}{n_{sym,0}} \approx \frac{1}{C\, \alpha^2} \frac{M_{\gamma_{DM}}^2 \, T_{D}^2}{m_{pl}\, m_{\chi}^3} \,.
\ee
If the decoupling temperature is say $m_{\chi}/10$ (just an example) with $m_{\chi} \sim 1 \tev$ and if $M_{\gamma_{DM}} \sim O(\tev)$ (see the section on direct detection), the density at decoupling would be $n_{sym,D} \sim 10^{-16} n_{sym,0} $. Since one expects $\Delta n_{\chi} \sim 10^{-9} n_{sum,0}$ and that 86 \% of the asymmetric part remains, one can see that the number density of symmetric DM at $U(1)_{DM}$ decoupling is negligibly small compared with the asymmetric relic density.

Needless to say, a detailed analysis of luminogenesis is indeed extremely important. This will be treated elsewhere. What has been presented here
could be considered to be the first steps of an extended program of luminogenesis.

\section{Dark matter hadrons and small scale structure problem}

In a subsequent paper \cite{kevin}, we will show that, starting from the $SU(2)_L$ coupling at the 
electroweak scale and running it toward the unification scale $M_{DUT}$, one can deduce the
value of the DM gauge coupling of $SU(4)$ at that scale. From hereon, we shall call $SU(4)$ by the name Dark QCD (
DQCD).
Running it backward, one can look for the energy scale
at which $\alpha_4 \sim 1$. This will be, to a good approximation, the scale where DQCD confinement occurs.
As we have discussed briefly in \cite{paulpqdm}, dark baryons are formed by a bound state of {\em four} $\chi$'s resulting in massive {\em bosons}.
For definiteness, we shall call these dark baryons by the name $\chi$ Massive Particle or CHIMP. To a first approximation, the CHIMPs
will have a mass of approximately {\em four} times the DQCD confinement scale $\Lambda_4$. 

As in \cite{paulpqdm}, there are three flavors
of dark matter fermion $\chi$ (one for each family). In the absence of explicit mass terms, there is a chiral symmetry $SU(3)_L \times SU(3)_R$ among $\chi$'s. From QCD (restricting oneself to two flavors for simplicity), we learn that $\langle \bar{q} q \rangle \neq 0$ spontaneously breaks the quark chiral symmetry resulting in the appearance of Nambu-Goldstone (NG) bosons which however acquire a small mass due to the explicit breaking of that chiral symmetry coming from the small masses of the up and down quarks and thus becoming what are called pseudo NG bosons. We expect a similar phenomenon to occur for DQCD. $\langle \bar{\chi} \chi \rangle \neq 0$ would yield {\em massless} NG bosons. It will be seen below that one has to break explicitly the DM chiral symmetry by a tiny amount in order to endow these NG bosons with a tiny mass. To be specific, the explicit breaking of the $\chi$-chiral symmetry can be parametrized by a term $m_0 \bar{\chi} \chi$ with $m_0 \ll \Lambda_4$. $m_0$ will be the free parameter of the model which will be determined by the fit to the small scale, Milky Way and possibly cluster scale anomalies as discussed below. For definiteness, we shall denote these pseudo-NG bosons as $\pi_{DM}$. The arguments go as follows.

Below $\Lambda_4$, one can write down an effective theory of CHIMPs interacting with the pseudo-NG bosons, very much in the same vein as nucleon-pion interactions with the difference being that in our case the CHIMPs are {\em bosons} instead of being fermions. In some sense, the dynamics of DQCD would presumably be simpler than that of QCD since CHIMPs carry no spin. One can write down a non-relativistic potential between two CHIMPs exchanging a $\pi_{DM}$ as
\be
\label{DMpotential}
V= - \frac{\alpha_{DM}}{r} \exp(-m_{\pi_{DM}} r) \, .
\ee
Such a potential has been investigated phenomenologically by \cite{zurek} although the Lagrangian is for a Yukawa interaction between a fermionic DM and a scalar. Non-relativistically it is the same. Here, we provide an explicit model for the spin-0 field, namely the pseudo-NG bosons- the dark pions- of DQCD. 

As an example (although a more detailed investigation is surely needed), one can use Fig. 6 of \cite{zurek} to get a very rough estimate of the parameter range allowed to solve the small scale structure anomalies. First, a word of caution is in order here. Our CHIMP-dark pion interactions are presumably strong judging from what we know about pion-nucleon interactions although QCD can be {\em very different} from DQCD. Results shown in Fig. 6 of \cite{zurek} are for perturbative values of $\alpha_{DM}$ up to $\alpha_{DM} =0.1$. For the sake of argument, let us take $\alpha_{DM} =0.1$ to make our estimate. Taking into account only small scale structure anomalies, one can extrapolate to see that masses of CHIMPs ranging up to 100 TeV or so and $m_{\pi_{DM}}< 1 MeV$. If one would like to accommodate also Milky Way and cluster bounds, the CHIMP mass is seen to be lower, in the range of a few hundreds of GeV to a few TeVs, with 
$m_{\pi_{DM}}> 1 MeV$. Notice again that the dark pion mass $m_{\pi_{DM}}$ is related $m_0$ which appears in the explicit breaking term of the $\chi$-chiral symmetry $m_0 \bar{\chi} \chi$ with $m_0$ being a free parameter.

It goes without saying that much remains to be done to tackle these issues in the framework of strong coupling regime as in our model. But it is encouraging that light pseudo scalars appear naturally due to the chiral symmetry of the model and this lightness appear to be what might be needed to solve the small scale structure anomalies and perhaps larger scales as well.

\section{Grand Unification Reconsidered}

Since 1974, a great deal of research has proceeded based on the 
idea that the SM gauge group is contained is a 
larger grand unified GUT
group $G_{GUT}$. The simplest GUT model is based
\cite{GeorgiGlashow} on $G_{GUT} \equiv SU(5)$ which, in its 
minimal form, makes a sharp prediction for the proton decay 
lifetime
based\cite{GeorgiQuinnWeinberg} on a GUT scale 
$M_{GUT} \gtrsim 10^{14}$ GeV. Experimental searches excluded this
prediction already in 1984 but many alternative GUT 
theories are viable which survive this test. 
Accurate unification of 
the SM couplings at $M_{GUT} \sim 2 \times 10^{16}$ GeV
has frequently been cited 
\cite{Amaldi1,Amaldi2} as evidence for supersymmetry, and
GUT theories are an intermediate goal in much of
string theory phenomenology.\\

\noindent
By contrast, in the present luminogenesis model there is no luminous matter
with mass above $M_{15} \sim 10^9$ GeV, so that extrapolation
of the SM gauge couplings to orders
of magnitude above the $T_{15}$ scale, while including only
luminous matter states in the calculation of the 
renormalization group flow, is rendered
physically inappropriate. This provides a plausible rationale for the
non-confirmation of the proton lifetime predicted on the basis of such
an extrapolation in e.g. $SU(5)$. In the present model based on the gauge group of Eq.~(\ref{group}), proton decay is absent.

\section{Direct detection}

The direct detection of dark matter in our model 
can come about by the exchange of the dark photon,
$\gamma_{DM}$. Dark matter can interact with luminous 
matter in the direct detection search through the 
exchange in the t-channel of the massive dark photon, 
namely through the use of Eq.~(\ref{gamma}). An estimate 
of the mass $M_{\gamma_{DM}}$ assuming $g = O(1)$ using the 
bound by XENON100 \cite{XENON100} for the cross section for 
a dark matter mass of e.g. 1 TeV, namely 
$\sigma < 10^{-44} cm^2$ gives $M_{\gamma_{DM}} > O(2 \, TeV)$. 
Nevertheless, we can eagerly await results from
the upgraded version of XENON100 to XENON1T
being planned \cite{XENON1T}
for direct detection of dark matter particles.

\section{Discussion}

Our principal underlying assumption is that in the very early
universe the inflaton decays into only dark matter
and that at a later, though still early, cosmological era,
luminogenesis converted some 14 \% of this
dark matter into luminous matter. Our specific model
gives rise naturally to strongly-interacting dark matter
which can overcome some important short-range problems
confronting cold dark matter. Luminogenesis occurs via
an extremely weak interaction characterized
by a mass scale $\sim 10^7-10^9 \gev$. The possible irrelevance 
of grand unified GUT models which include 
only luminous matter is clarified
in this broader perspective. Finally, higher sensitivity
direct detection of dark matter will be of crucial importance
in sharpening our understanding of the
luminogenesis stage in the early universe.
   
\section{Acknowledgments}
PHF was supported in part by US DOE grant DE-FG02-06ER41418.
PQH was supported in part by US DOE grant DE-FG02-97ER41027.
PQH would like to thank Alexander Kusenko for the stimulating atmosphere at PACIFIC 2013 where this work was completed.

\end{document}